\documentclass[twocolumn, journal]{IEEEtran} 

\makeatletter 
\def\ps@headings{%
\def\@oddhead{\mbox{}\scriptsize\rightmark \hfil \thepage}%
\def\@evenhead{\scriptsize\thepage \hfil \leftmark\mbox{}}%
\def\@oddfoot{}%
\def\@evenfoot{}} 
\makeatother 
\usepackage{latexsym,times,epsf,amsmath,amssymb,amsfonts,graphicx}
\usepackage{subfigure}
\usepackage{multirow}
\usepackage{algpseudocode}
\usepackage{algorithm}
\usepackage{mathrsfs}
\usepackage{array}
\usepackage{setspace}
\usepackage{cite}
\usepackage{csquotes}
\usepackage{comment}
\usepackage{subfigure}
\usepackage{cite}
\usepackage{amsthm}
\usepackage{amsmath}

\usepackage{color}
\usepackage{bbm}

\renewcommand{\theequation}{\arabic{equation}}
\newtheorem{Theorem}{\hskip 0pt Theorem}
\newtheorem{Lemma}{\hskip 0pt Lemma}

\allowdisplaybreaks
\begin{document}

\title{On Channel State Feedback Model and Overhead in Theoretical and Practical Views}


\author{Chunbo Luo, University of Exeter, c.luo$@$exeter.ac.uk}



\maketitle

\begin{abstract}

Channel state feedback plays an important role to the improvement of link performance in current wireless communication systems, and even more in the next generation. The feedback information, however, consumes the uplink bandwidth and thus generates overhead. In this paper, we investigate the impact of channel state feedback and propose an improved scheme to reduce the overhead in practical communication systems. Compared with existing schemes, we introduce a more accurate channel model to describe practical wireless channels and obtain the theoretical lower bounds of overhead for the periodical and aperiodical feedback schemes. The obtained theoretical results provide us the guidance to optimise the design of feedback systems, such as the number of bits used for quantizing channel states. We thus propose a practical feedback scheme that achieves low overhead and improved performance over currently widely used schemes such as zero holding. Simulation experiments confirm its advantages and suggest its potentially wide applications in the next generation of wireless systems.

\end{abstract}


\section{Introduction}
\label{sec:introduction}

One of the key technologies in 4G communication systems is channel state information (CSI) feedback, which is gaining increased importance in the research and design of 5G systems that are undergoing the standardization stage. The importance will be even significant in 6G which includes communication and satellite positioning system~\cite{Huang2014, Huang2011,Huang2011a,Huang2013b}. The importance of CSI feedback can be simply summarized as that more knowledge of the wireless channel can enhance the system's throughput. Such advantage is achieved through novel schemes including precoding, modulation and coding scheme (MCS) selection, or beam selection. As an essential technology for the 5G systems,  millimeter wave provides exceptionally high throughput, but also comes with the down sides such as reflection, scattering, and blockage, which cause more frequent change of channel states since its wave length is shorter than traditional wireless carriers. Therefore, the feedback of CSI needs to be more frequent than current cellular systems in order to maintain a high throughput link. The increased feedback data would generate a lot of overhead and consume excessive bandwidth that is essential for other types of services in 5G systems. To optimize the feedback overhead is thus a key research topic for improving spectrum resource, which motivates this paper.

 


The benefits of CSI feedback have been recognized for long time and exploited in contempoary wireless systems. We will briefly introduce several recently reported examples. In~\cite{Lakshmana2016, Anand2017}, the channel states sent back to a transmitter are used for precoding. In~\cite{Wang2016, Rezaee2016}, authors proposed novel interference analysis and cancellation schemes driven by the channel states feedback information. Authors of \cite{Abedi2016, Javan2016} studied the benefits of CSI feedback for optimizing resource allocation. In addition to these benefits, pioneering studies on physical layer security have showed the power of CSI feedback on the improvement of communication security~\cite{Yang2016, Wang2016a}.

The benefits of CSI feedback come with the cost of increased overhead that consumes the uplink spectrum resource. For the purpose of reducing this cost, optimization schemes have been widely studied. In~\cite{Eltayeb2014,Lv2016}, compressed sensing~(CS) is leveraged to reduce the number of the bits for CSI feedback. Two issues are in association with the CS based method. First, the delay is significantly large since CS is a blockwise signal processing method and is usually solved via a large number of iterative calculation. Second, CS relies on the presumption of the downlink channel being sparse. Such a sparsity assumption has not been fully investigated in research. As a more matured method, a discrete cosine transform~(DCT) basis matrix is designed to compress the channel states in~\cite{Kuo2012}. DCT is a blockwise signal processing method, and the efficiency is proportionally related to the length of the data blocks for DCT. Therefore, there exists a trade-off between the compression efficiency and timeliness of CSI feedback.




As shown in the generation of fading channels~\cite{Huang2013,Huang2013a,Huang2016}, fading channel is indeed a stationary random process which allows us to predict a channel state based on its  previous several observations. This model is also used in the estimation of the channel state information~\cite{Huang2013c}. The difference of a prediction versus its real channel state is usually called as the innovation of a channel. For a stationary fading channel, the feedback of innovation can be utilized to reconstruct full channel states. Generally, the quantization of the innovation needs less bits than the quantization of original channel. Therefore, the feedback of quantized innovation is also recognized as an effective CSI feedback technology. In literature, the quantization and feedback of channel innovation are usually called as differential channel state feedback methods. 

The order-one autoregressive~(AR) model is often used to model a fading channel \cite{Zhou2013, Jeon2016}. Based on the AR(1) model, the difference between a channel state $x_k$ and $\hat{x}_k$, where the latter is the prediction of $x_k$ based on $x_{k-1}$, is quantized. In such a differential feedback scheme, the minimum number of bits, known as feedback overhead, is estimated. Similar to the method in~\cite{Zhou2013}, AR(1) model based differential quantization is also investigated in~\cite{Jeon2016}. Different from the scalar quantization in~\cite{Zhou2013}, vector quantization is considered in~\cite{Jeon2016}, which claims that vector quantization has advantages over the scalar one. 
On the transmission of CSI feedback, several methods have been proposed. In~\cite{Kim2012}, there is no feedback loop at the CSI transmitter side, and quantization noise will be accumulated when the channel state is reconstructed at the receiver side. The same problem also exists in the method proposed in~\cite{Zhang2012}. Different from the work in~\cite{Kim2012}, rate distortion theory is employed to estimate the lower bound of the quantization bits under the constraint of channel reconstruction mean square error (MSE)~\cite{Zhang2012}. However, we can deduce that the rate distortion function calculated in~\cite{Zhang2012} does not have the right distortion when the number of quantization bits is zero.  

Based on the introduction above, there are basically two types of CSI feedback schemes. The first type focuses on the feedback of original channel states~\cite{Eltayeb2014,Lv2016,Kuo2012}, and the second one is based on the transmission of differential channel states \cite{Jeon2016,Zhou2013,Kim2012,Zhang2012,Nevat2016,Szurley2017}. The latter ones are gaining more popularity because of their ability to reduce redundancy between adjacent channel states~\cite{Jake1974}, and thus require less data bits than the schemes that transmit the original channel states. However, existing differential feedback schemes are of problems regarding the following three aspects. First, the channel modelling is not accurate. In literature, the order one autoregressive structure such as the AR(1) model is widely used to describe a fading channel, and is not accurate description of Rayleigh or Rician channels. Furthermore, channel fading and additive noise are not simultaneously considered in literature, i.e., the additive white Gaussian noise (AWGN) is not taken into account at the modelling of channel fading. Third, open loop differential schemes are considered in literature, but they have the high risk of accumulating quantization noise. A closed loop structure is necessary to correct the channel distortion in time and avoid noise accumulation.





In this paper, we consider fading channels to be the response of white Gaussian variables passing through an autoregressive structure. Different from the AR(1) model, the order of the autoregressive structure is not assumed to be in rank one, rather a more realistic setting greater than one and extensible to infinitely large. Besides, AWGN is simultaneously considered with channel fading in our analysis. Based on the more complete channel model, we first investigate the effect of operation delay on the overhead. To reduce the accuracy degradation caused by delay, we predict the channel state given its previous a few observations. Since the prediction has intrinsic finite accuracy, we use the channel state mean square error~(MSE) to investigate the accuracy, which is essentially equal to the channel reconstruction distortion in the extreme case of infinite-bits feedback. Afterwards, we obtain the theoretic lower bound of MSE given finite-bits of CSI feedback. Both periodical and aperiodical feedback schemes adopted by standards are analysed using the rate distortion theory. Under the guidance of theoretic results, we propose a novel and practical CSI feedback method based on more realistic channel modelling and parameter configuration, which achieves improved long-term performance over contemporary methods, confirmed by both theoretical study and extensive experiments. 

The rate distortion theory based feedback overhead optimization is also performed in~\cite{Huang2017} where the theoretic work is performed. We pay more attention to finding a practical solution to optimizing feedback overhead. The network overhead for compressed sensing is estimated in~\cite{Huang2014a, Huang2014b,Huang2016,Huang2017a}.

The rest of this paper is organized as follows. Introduction to traditional channel feedback systems and problem formulation are presented in Section~\ref{sec: problem form}. In Section~\ref{sec: accuracy comparison}, we analyze the minimum MSE of the extreme case of infinite bits used for channel state feedback, and compare our results with that of channel estimation based on the zero-holding rule. The rate distortion theory is employed to calculate the theoretic lower bounds on bits for both aperiodical and periodical feedback schemes in~\ref{sec: theoretic bounds}. We propose a novel channel state feedback scheme with performance analysis in Section~\ref{sec: a KF compression}. Numerical simulations are performed in Section~\ref{sec: simulations} followed by conclusions in Section~\ref{sec: conclusion}. 



\section{Problem Formulation}
\label{sec: problem form}

\begin{figure}[t]
\centering
\includegraphics[width=1\linewidth] {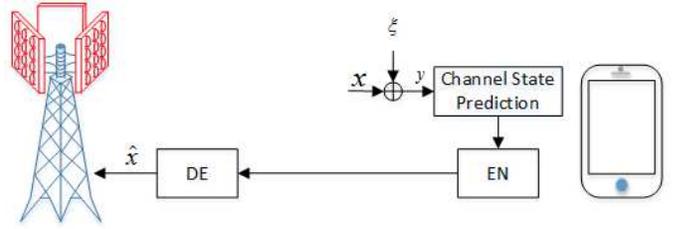}
\setlength{\abovecaptionskip}{3pt plus 3pt minus 2pt}
\caption{System diagram.}
\label{fig:system}
\end{figure}

The channel state feedback scheme is shown in Fig.~\ref{fig:system}. The fading channel is denoted by $x$. Let $x_k$ denote the channel sample at the $k$-th time instance. The fading channel state is not accessible due to the AWGN which is denoted by $\xi$. The additive noise at the $k$-th time instance is $\xi_k$. The overall degraded channel, accessible from the measurements of the reference signals, can be denoted as follows,
\begin{equation}
y_k=x_k+\xi_k.
\label{eq: 1 observation}
\end{equation}

The widely adopted Rayleigh channel model is also used in our analysis, which can be extended to other channel models following a similar method. According to the definition of Rayleigh distribution, both the In-phase and Quadrature component of $x$ follow a Gaussian distribution. It is easy to know that a Gaussian random process is the response of a white Gaussian variable passing through an AR structure that determines the spectrum of the random process.  Let $\psi_k$ denote this variable, and $x_k$ can be modelled as follows
\begin{equation}
x_k=\sum_{m=1}^L a_m x_{k-m} +\psi_k,
\label{eq: chan math model}
\end{equation}
where $\psi_k$ is a variable following the Gaussian distribution with mean zero and variance of $\sigma_{\psi}^2$, and $L$ is the memory depth of the random process equal to the order of the AR structure. 

Due to the protocol stack and signal processing procedure, there inevitably exists operation delay in a feedback system. To counteract the accuracy degradation caused by delays, a series of noisy observations of the fading channel are used to obtain one-step forward predictions of the channel state. Without loss of generality, the noisy observations are represented by $Y_{k,1:L}=(y_{k-L+1}, y_{k-L+2},\cdots,y_{k})^T$. 

With the predicted channel state or the channel innovation, the user equipment (UE) will encode and send it to the base station. The transmission of the codewords consumes the uplink bandwidth. To quantitatively evaluate the performance of a feedback system, we calculate the average bit rate $R$ required for sending through the codewords. Details about $R$ and the encoding procedure will be introduced in the following sections. At the base station, the codewords are decoded to reconstruct the fading channel, denoted by $\hat{x}_{k+1}$.

As in the introduction of the system model, only the noisy observation of channel $y$ can be directly read, rather than the original channel state $x$. $R$ is thus related to $y$. Different concrete quantization methods generate different size of $R$. To remove the limits brought in by concrete methods, the mutual information between $y$ and $\hat{x}$ is taken as the metric to express $R$. Essentially, the mutual information is a theoretical lower bound on $R$ which is written as follows,
\begin{equation}
I(y_{k+1};\hat{x}_{k+1}).
\label{eq: x hat x MI}
\end{equation}

Furthermore, we can straightforwardly investigate the tradeoff between $R$ and the channel reconstruction accuracy. Generally, higher accuracy requires more bits for quantization. The accuracy is measured by  MSE between the real channel state $x_{k+1}$ and its reconstruction $\hat{x}_{k+1}$, which is defined as follows,
\begin{equation}
D_x=E\left[\left(x_{k+1}-\hat{x}_{k+1}\right)^2\right],
\label{eq: x MSE}
\end{equation}
where $E[\cdot]$ denotes the expectation. 

We can now formulate the optimization problem that aims to minimize the number of bits under the constraint of channel reconstruction MSE,
\begin{equation}
\left\{\begin{array}{rl}
\mbox{Objective}: & R=\inf\left\{\frac{1}{L}I(y_{k,1:L}; \hat{x}_{k+1})\right\}\\
\mbox{subject to}: & d(x_{k+1}, \hat{x}_{k+1})\leq D_x
\end{array}
\right..
\label{eq: opt original def1}
\end{equation}

For the purpose of analysis, we introduce an auxiliary variable $\hat{y}_{k+1}$. The physical sense of $\hat{y}_{k+1}$ is regarded as the reconstructed $y_{k+1}$. The relationship between $\hat{y}_{k+1}$ and $\hat{x}_{k+1}$ is defined as follows,
\begin{equation}
\hat{x}_{k+1}\overset{\Delta}{=}\frac{\sigma_x^2}{\sigma_x^2+\sigma_{\xi}^2}\hat{y}_{k+1}.
\label{eq: auxiliary y}
\end{equation}

Since $\hat{x}_{k+1}$ is the scaled $\hat{y}_{k+1}$ by a constant, the following two definitions of mutual information are equal to each other: $\frac{1}{L}I(y_{k,1:L}; \hat{x}_{k+1})=\frac{1}{L}I(y_{k,1:L}; \hat{y}_{k+1})$. The optimization problem defined in (\ref{eq: opt original def1}) is thus converted to the following form: 
\begin{equation}
\left\{\begin{array}{rl}
\mbox{Objective}:& R=\inf\left\{\frac{1}{L}I(y_{k,1:L}; \hat{y}_{k,1:L})\right\}\\
\mbox{subject to}: & d(x_{k+1}, \hat{x}_{k+1})\leq D_x
\end{array}
\right..
\label{eq: opt original def2}
\end{equation}

As designed in the 3GPP standards, channel states can be sent back to the base station periodically or aperiodically. The differences between these two methods have fundamental impact towards the solution of (\ref{eq: opt original def2}). The following sections will investigate both cases in order to find the optimal feedback schemes under the constraint of MSE.


\section{An Extreme Case: Infinite Bits Feedback and Reconstruction MSE}
\label{sec: accuracy comparison}


The previous section introduces two metrics to evaluate a channel state feedback scheme: channel reconstruction accuracy and the number of bits. In this section, we consider that channel states are sent to BS in real numbers, which can only be achieved by representing a channel state with infinite bits. The work in this section thus reveals the reconstruction distortion $D_x$ at the extreme case of $R\rightarrow\infty$.  

In literature, the channel state observed at the current time instance is usually adopted as the state at the next time instance. Such a channel state determination strategy is called \textit{zero-holding}~(ZH). We will compare the MSE of the reconstructed channel states from the ZH strategy and that from our prediction method. 



\subsection{MMSE in one-step ahead prediction based on real-valued channel states feedback}


Since infinite large number of bits are used to quantize the channel state feedback, there will be no quantization loss. The channel reconstruction accuracy is fully determined by the error occurring in the one-step ahead prediction of $x_{k+1}$ given $y_{k,1:L}$. Thus, the lower bound on the distortion $D_x$ at $R\rightarrow\infty$ is equal to the MMSE of $x_{k+1}$,
\begin{equation}
D_x|_{R\rightarrow\infty}=E\left[\left(x_{k+1}-x_{k+1}^{\prime}\right)^2\right],
\label{eq: mmse pred}
\end{equation}
where $x_{k+1}^{\prime}$ is the MMSE prediction of $x_{k+1}$ given the real-valued channel states feedback, that is,
\begin{equation}
{x}_{k+1}^{\prime}=E[x_{k+1}|y_{k,1:L}].
\label{eq: predict model}
\end{equation}

Since $y_{k}$, $k\in\mathbb{Z}$, is the noisy observation of $x_{k}$ under additive Gaussian noise $\xi_{k}$, the MMSE estimation of $x_{k}$ given $y_k$ is in a linear form. Furthermore, we can easily prove that $y$ is a spherical invariant random process. Thus, there exists a linear extrapolation of $y_{k+1}$ based on $y_{k,1:L}$. Furthermore, we know that there exists a MMSE linear prediction of $x_{k+1}$ given $y_{k,1:L}$. Let $\Theta$ denote the coefficients of the linear predictor. Based on the results in~\cite{Sayed2003}, the coefficient vector $\Theta$ of the MMSE predictor is calculated by
\begin{equation}
\begin{aligned}
\Theta=\left[\begin{array}{c} \kappa_{xy}(1) \\ \kappa_{xy}(2) \\ \vdots \\ \kappa_{xy}(L) \end{array}\right]^{T}\bold{K} 
\end{aligned}
\label{eq: MMSE prediction coef}
\end{equation}
where $\kappa_{xy}(i)=E[x_{k+1}y_{y+1-l}]$, $l\in\mathcal{L}$ and $\mathcal{L}=\{1,2,\cdots,L\}$ and 
\begin{equation}
\begin{aligned}
\bold{K}=\begin{bmatrix}
\kappa_y(0) & \kappa_y(1) & \cdots & \kappa_y(L-1)\\
\kappa_y(1) & \kappa_y(0) & \cdots & \kappa_y(L-2) \\
\vdots & \vdots & \ddots &  \vdots\\
\kappa_y(L-1) & \kappa_y(L-2) & \cdots & \kappa_y(0)
\end{bmatrix}.
\end{aligned}
\label{eq: kappa matrix}
\end{equation}

With the defined order-$L$ coefficients $\Theta$, the MMSE prediction of $x_{k+1}$ given $y_{k,1:L}$ is calculated as follows,
\begin{equation}
x^{\prime}_{k+1}=\Theta\cdot
\begin{bmatrix}y_{k-L+1}\\
y_{k-L+2}\\
\vdots\\
y_{k}
\end{bmatrix}.
\label{eq: linear prediction}
\end{equation}


According to the results in~\cite{Sayed2003}, the mean square error in real-valued channel state feedback $D_x|_{R\rightarrow\infty}$ is calculated via
\begin{equation}
\begin{aligned}
D_x|_{R\rightarrow\infty}&=\sigma_x^2-\left[\begin{array}{c} \kappa_{xy}(1) \\ \kappa_{xy}(2) \\ \vdots \\ \kappa_{xy}(L) \end{array}\right]^{\prime}\cdot\bold{K}^{-1}\cdot\left[\begin{array}{c} \kappa_{xy}(1) \\ \kappa_{xy}(2) \\ \vdots \\ \kappa_{xy}(L) \end{array}\right]\\
&\overset{\Delta}{=}\sigma_P^2.  
\end{aligned}
\label{eq: MMSE prediction e}
\end{equation}

The distortion of the reconstructed channel based on one-step ahead prediction has been obtained for the extreme case of $R\rightarrow\infty$. Next, the distortion for ZH strategy will be calculated. 

\subsection{MMSE in zero-holding channel state determination}
\label{subsec: zero holding mse}

In the literature, not much attention has been paid on the operation delay of channel state feedback. The current channel state is taken as the one at the next time instance which is called as zero-holding. In this subsection, we will quantitatively determine the distortion occurring in the zero-holding strategy as a contrast to the one-step ahead prediction strategy. Let $\hat{x}_{k+1}^Z$ denote the channel state at $(k+1)$-th time instance determined via the ZH rule. The ZH strategy can be mathematically described as follows,
\begin{equation}
\hat{x}_{k+1}^Z=x_{k}^{\dagger},
\label{eq: zero hold est}
\end{equation}
where $x_{k}^{\dagger}$ denotes the estimation of $x_k$.

The MMSE estimation of current channel state is an optimum estimation given additive Gaussian noise, which is widely used in literature. Since $x_{k}^{\dagger}$ is the MMSE estimation of $x_k$ given $y_k$, we have the following equation,
\begin{equation}
x_{k}^{\dagger}=E\left[x_k|y_k\right].
\label{eq: x k mmse}
\end{equation}

Channel distortion is similarly measured by the MSE of the reconstructed channel $x_{}^{\dagger}$ versus its real value $x$ as follows, 
\begin{equation}
\begin{aligned}
\sigma^2_Z
=&E\left[\left(x_{k+1}-\hat{x}_{k+1}^Z\right)^2\right]
=E\left[\left(x_{k+1}-x_{k}^{\dagger}\right)^2\right]\\
=&E\left[\left(x_{k+1}-x_{k}-(x_k-x_{k}^{\dagger})\right)^2\right]\\
=&E\left[\left(x_{k+1}^2\right)\right]
+E\left[\left(x_{k}^2\right)\right]
+E\left[\left((x_k-x_{k}^{\dagger})^2\right)\right]\\
-&2E\left[\left(x_{k+1}x_{k}\right)\right]
-2E\left[\left(x_{k+1}(x_k-x_{k}^{\dagger})\right)\right]\\
+&2E\left[\left(x_{k}(x_k-x_{k}^{\dagger})\right)\right]\\
\overset{(a)}{=}&2\sigma_x^2-2\kappa_x(1)+\frac{\sigma_x^2\sigma_{\xi}^2}{\sigma_x^2+\sigma_{\xi}^2},
\end{aligned}
\label{eq: mse zero hold}
\end{equation}
where $\sigma^2_Z$ denotes the MSE, and in $(a)$, the first two terms are calculated as follows
\begin{equation}
\begin{aligned}
&E\left[\left(x_{k+1}^2\right)\right]=E\left[\left(x_{k}^2\right)\right]=\sigma_x^2\\
&E\left[\left(x_{k+1}x_{k}\right)\right]=\kappa_x(1)\\
\end{aligned}
\end{equation}
following the definition of autocorrelation of a random process. The third term in $(a)$ is calculated as follows
\begin{equation}
\begin{aligned}
&E\left[\left((x_k-x_{k}^{\dagger})^2\right)\right]=\frac{\sigma_x^2\sigma_{\xi}^2}{\sigma_x^2+\sigma_{\xi}^2},
\end{aligned}
\end{equation}
where the orthogonal principle has been employed,
\begin{equation}
\begin{aligned}
E\left[\left(x_{k}(x_k-x_{k}^{\dagger})\right)\right]=E\left[\left(x_{k+1}(x_k-x_{k}^{\dagger})\right)\right]=0.
\end{aligned}
\end{equation}

We have obtained the channel reconstruction distortion for both the one-step ahead prediction scheme and ZH strategy. The results are obtained at the extreme case that infinite number of bits are used to quantize the channel state. Next, we will investigate a more practical case that only finite bits are used to convey the channel states from a UE to a base station.

\section{Theoretic Lower Bounds on Finite-Bits Represented Channel States Feedback}
\label{sec: theoretic bounds}



As mentioned in Section~\ref{sec: problem form}, the two types of channel state feedback schemes in 3GPP LTE, aperiodic and periodic feedback transmission, are also considered in the next generation of cellular systems~(5G). It is thus significantly important to know the theoretical lower bounds of the length of bits required by these two schemes. 


\subsection{An Overhead Lower Bound in Aperiodic Feedback}
\label{subsec: aperiodic overhead}

The aperiodic feedback scheme works in such a protocol that a base station sends a request of downlink channel states when needed, and then the UE sends the channel state information back to the base station, which is unavoidably delayed due to the protocol stack and signal processing. This reactive style implies that the time interval between two channel state transmissions keeps changing. Since the feedback intervals vary, the vector quantization does not fit for compressing the channel states. Therefore, channel states are quantized via sample-wise quantization, and sent back to the base station. We can thus obtain the lower bound of the scalar quantization method, which is related to the entropy of a single channel sample as follows.



\subsubsection{Aperiodic feedback of one-step ahead predicted channel state}

The direct solution to (\ref{eq: opt original def2}) can hardly be achieved. Before calculating the theoretic lower bound, we define an auxiliary variable $d_y$, the MSE between $y_{k,1:L}$ and $\hat{y}_{k,1:L}$, as follows,
\begin{equation}
d_y=\sum_{l=1}^L E[\theta_l^2\left(y_{k-l+1}-\hat{y}_{k-l+1}\right)^2],
\label{eq: Dy def}
\end{equation}
where $\theta_l$ is an element of the matrix $\Theta$, $\Theta=\{\theta_l\}$, $l\in\mathcal{L}$ defined in (\ref{eq: MMSE prediction coef}).


With the auxiliary variable $d_y$, the auxiliary optimization problem can be formulated as follows,
\begin{equation}
R(D_y)=\inf_{d_y\leq D_y} \frac{1}{L}I(y_{k,1:L}; \hat{y}_{k,1:L}).
\label{eq: auxiliary RD aperiod}
\end{equation}


According to the results in~\cite{Berger1975}, the solution to (\ref{eq: auxiliary RD aperiod}) can be directly obtained:
\begin{equation}
R(D_y)=\frac{1}{2L}h(y_{k,1:L})-\frac{1}{2}\log2\pi e D_y +\frac{1}{2}\sum_{l=1}^L\log\theta_l^2.\\
\label{eq: auxiliary RD aperiod solu}
\end{equation}

For the aperiodic finite-bits feedback scheme, the MSE between $x_{k+1}$ and its reconstructed version at the base station can be calculated as follows,
\begin{equation}
\begin{aligned}
d_x^{AP}&=E\left[\left(x_{k+1}-\hat{x}_{k+1}\right)^2]\right]\\
&=E\left[\left(x_{k+1}-x^{\prime}_{k+1}+x^{\prime}_{k+1}-\hat{x}_{k+1}\right)^2]\right]\\
&=E\left[\left(x_{k+1}-x^{\prime}_{k+1}\right)^2]\right]+E\left[\left(x^{\prime}_{k+1}-\hat{x}_{k+1}\right)^2]\right]\\
&+2E\left[\left(x_{k+1}-x^{\prime}_{k+1}\right)\left(x^{\prime}_{k+1}-\hat{x}_{k+1}\right)]\right]\\
&\overset{(a)}{=}\sigma_P^2+E\left[\left(\sum_{l=1}^L \theta_l y_{k+1-l}-\sum_{l=1}^L \theta_l\hat{y}_{k+1-l}\right)^2\right]\\
&+2E\left[\left(x_{k+1}-x^{\prime}_{k+1}\right)\left(x^{\prime}_{k+1}-\hat{x}_{k+1}\right)\right]\\
&\overset{(b)}{=}\sigma_P^2+E\left[\left(\sum_{l=1}^L \theta_l y_{k+1-l}-\sum_{l=1}^L \theta_l \hat{y}_{k+1-l}\right)^2\right]\\
&\overset{(c)}{=}\sigma_P^2+\sum_{l=1}^L E\left[\theta_l^2 \left( y_{k+1-l}-\hat{y}_{k+1-l}\right)^2\right],\\
\end{aligned}
\label{eq: dist connection}
\end{equation}
where $(a)$ follows (\ref{eq: MMSE prediction e}); $(b)$ follows the orthogonal principle, and the quantization noise being an independent and zero mean variable; $(c)$ follows the derivation in the \textit{test} channel shown in~\cite{Berger1971}.


Combining (\ref{eq: auxiliary RD aperiod solu}) and (\ref{eq: dist connection}), we obtain the lower bound on the feedback overhead:
\begin{equation}
\begin{aligned}
&R(D_x^{AP})
=\frac{1}{2L}h(y_{k,1:L})-\frac{1}{2}\log2\pi e \left(D_x-\sigma_P^2\right) +\frac{1}{2}\sum_{l=1}^L\log\theta_l^2\\
&=h(y_{k,1:L})-\frac{1}{2}\log2\pi e \left(D_x-\sigma_P^2\right) +\frac{1}{2}\sum_{l=1}^L\log\theta_l^2\\
&\overset{(a)}{=}\log|\bold{K}_y|-\frac{1}{2}\log \left(D_x-\sigma_P^2\right)+\frac{1}{2}\sum_{l=1}^L\log\theta_l^2 \\
\end{aligned}
\label{eq: RD aperiod solu}
\end{equation}
where $\bold{K}_y=\{\kappa_y(i,j)=\kappa_y(i-j)\}$, and $(a)$ follows 
\begin{equation}
\frac{1}{L}h(y_{k,1:L})=\frac{1}{2}\log2\pi e + \frac{1}{2L}\log|\bold{K}_y|
\label{eq: y entropy}
\end{equation}


\subsubsection{Aperiodic feedback in the zero-holding strategy}
\label{subsec: bound zero holding}

As introduced in the previous section, the ZH assumption is widely adopted in current feedback scheme. Remember that $\hat{x}^Z_{k+1}$ denotes the zero-holding channel state at the $(k+1)$-th time instance. The estimation of $x_k$ from its noisy observation $y_k$ is taken as the value of $\hat{x}^Z_{k+1}$, i.e., $\hat{x}^Z_{k+1}=x_k^{\dagger}=E[x_k|y_k]$.

To calculate the theoretic lower bound on overheads, we define an auxiliary variable $d^Z_y$ which is the weighted MSE between $y_{k}$ and ${y}^{\dagger}_{k}$,
\begin{equation}
d^Z_y=\alpha E\left[\left(y_{k}-{y}^{\dagger}_{k}\right)^2\right],
\label{eq: ZH Dy def}
\end{equation}
where ${x}^{\dagger}_{k}=\alpha {y}^{\dagger}_{k}$, and $\alpha=\frac{\sigma_x^2}{\sigma_x^2+\sigma_{\xi}^2}$ which is derived from the MMSE estimation of $x$ from its observation $y$ in AWGN.

MSE of the ZH scheme's $x_{k+1}$ and $\hat{x}^Z_{k+1}$ is calculated as follows,
\begin{equation}
\begin{aligned}
d^Z_x&=E\left[\left(x_{k+1}-\hat{x}^Z_{k+1}\right)^2\right]\\
&=E\left[\left(x_{k+1}-\alpha y_k+\alpha y_k -\hat{x}^Z_{k+1}\right)^2\right]\\
&\overset{(a)}{=}E\left[\left(x_{k+1}-\alpha y_k+\alpha y_k -\alpha {y}^{\dagger}_{k}\right)^2\right]\\
&=E\left[\left(x_{k+1}-\alpha y_k\right)^2\right]\\
&+E\left[\left(\alpha y_k -\alpha {y}_k\right)^2\right]\\
&+E\left[\left(x_{k+1}-\alpha y_k\right)\left(\alpha y_k -\alpha {y}^{\dagger}_{k}\right)\right]\\
&\overset{(b)}{=}\sigma_Z^2+d^Z_y
+E\left[\left(x_{k+1}-\alpha y_k\right)\left(\alpha y_k -\alpha \hat{y}_k\right)\right]\\
\end{aligned}
\label{eq: ZH x distortion}
\end{equation}
where $(a)$ follows (\ref{eq: zero hold est}), and $(b)$ follows (\ref{eq: mse zero hold}) and (\ref{eq: ZH Dy def}).

Different from the one-step prediction based on multiple previous observations, $y_{k,1:L}$, ${x}^{\dagger}_{k}$ is derived from a single observation $y_k$ in the ZH scheme. We can thus create an auxiliary optimization problem as follows,
\begin{equation}
R(D^Z_y)=\inf_{d^Z_y\leq D^Z_y} I(y_{k}; {y}^{\dagger}_{k}).
\label{eq: ZH auxiliary RD aperiod}
\end{equation}

The solution to (\ref{eq: ZH auxiliary RD aperiod}) is given as follows,
\begin{equation}
R(D_y^Z)=\frac{1}{2}h(y_{k})-\frac{1}{2}\log2\pi e D^Z_y +\frac{1}{2}\log\alpha.\\
\label{eq: ZH auxiliary RD aperiod solu}
\end{equation}


Given (\ref{eq: ZH x distortion}), we can derive the rate distortion function (\ref{eq: ZH auxiliary RD aperiod solu}) as follows,
\begin{equation}
\begin{aligned}
&R(D_y^Z)=\frac{1}{2}h(y_{k})+\frac{1}{2}\log\left(\frac{\sigma_{x}^2}{\sigma_{x}^2+\sigma_{\xi}^2}\right)\\
&-\frac{1}{2}\log2\pi e \left(\begin{matrix}D_x^Z-\sigma_z^2\\-E\left[\left(x_{k+1}-\alpha y_k\right)\left(\alpha y_k -\alpha \hat{y}_k\right)\right]\end{matrix}\right)\\
&\overset{(a)}{=}\frac{1}{2}h(y_{k})+\frac{1}{2}\log\left(\frac{\sigma_{x}^2}{\sigma_{x}^2+\sigma_{\xi}^2}\right)-\frac{1}{2}\log2\pi e \left(D_x^Z-\sigma_z^2\right),
\label{eq: ZH RDx aperiod solu}
\end{aligned}
\end{equation}
where $(a)$ is obtained according to the test channel, i.e,  $y=\hat{y}+\varkappa$. The test channel is the sufficient and necessary condition in which the equality holds in (\ref{eq: ZH auxiliary RD aperiod}). In the test channel, $\varkappa$ follows the Gaussian distribution with zero mean and variance of $D_y^Z$. Thus, we have
\begin{equation}
E\left[\left(x_{k+1}-\alpha y_k\right)\left(\alpha y_k -\alpha {y}^{\dagger}_{k}\right)\right]=0.
\label{eq: 0 crossing term}
\end{equation}

The theoretic lower bounds of the two aperiodic channel state feedback schemes have been obtained. The next subsection will investigate the theoretic bounds of the periodic feedback schemes.

\subsection{Periodic Feedback Overhead}
\label{subsec: periodic overhead}

Fading channels are usually a stationary random process, which implies that channel states nearby in time domain are correlated to each other. It is thus possible to only transmit different information (i.e., innovation) between samples and reduce the number of feedback bits. Innovation based transmission can be achieved through periodic feedbacks.




In order to calculate the required bits in the periodic feedback, we need to write the noisy observation $y$ with an autoregressive structure. As shown in (\ref{eq: chan math model}), a fading channel $x$ is the output of a white Gaussian variable $\psi$ passing through an autoregressive system. At the UE, observed channel states are essentially equal to the summation of $x$ and $\xi$, shown in (\ref{eq: 1 observation}). Combining (\ref{eq: 1 observation}) and (\ref{eq: chan math model}), we derive the autoregressive structure of $y$ as follows,
\begin{equation}
\begin{aligned}
y_k&=\sum_{m=1}^L a_m x_{k-m}+\psi_k+\xi_k\\
&\overset{(a)}{=}\sum_{m=1}^L a_m\left( \frac{\sigma_x^2}{\sigma_x^2+\sigma_{\xi}^2}y_{k-m}+e_{k-m}\right)+\psi_k+\xi_k\\
&=\sum_{m=1}^L a_m \frac{\sigma_x^2}{\sigma_x^2+\sigma_{\xi}^2}y_{k-m} +\sum_{m=1}^L a_m e_{k-m}+\psi_k+\xi_k\\
&\overset{(b)}{=}\sum_{m=1}^L a_m \frac{\sigma_x^2}{\sigma_x^2+\sigma_{\xi}^2}y_{k-m} +\nu_k,
\label{eq: y AR}
\end{aligned}
\end{equation}
where $e_{k-m}$ is the difference between an estimated channel state and its real value. $e_{k-m}$ observes zero mean Gaussian distribution with the variance of $\frac{\sigma_x^2\sigma_{\xi}^2}{\sigma_x^2+\sigma_{\xi}^2}$; $(a)$ follows the MMSE estimation of $x$ under the noise $\gamma$,
\begin{equation}
E[x|y]=\frac{\sigma_x^2}{\sigma_x^2+\sigma_{\xi}^2}y
\label{eq: mmse est x xi}
\end{equation}
$(b)$ follows the definition:
\begin{equation}
\nu_k\overset{\bigtriangleup}{=}\sum_{m=1}^L a_m e_{k-m}+\psi_k+\xi_k.
\label{eq: nu define}
\end{equation}

From the analysis above, $y$ is essentially equal to output of the \textit{i.i.d.} white Gaussian variable $\nu_k$, $\nu_k\thicksim\mathcal{N}(0,\sigma_{\nu}^2)$, $\sigma_{\nu}^2=\frac{\sigma_x^2\sigma_{\xi}^2}{\sigma_x^2+\sigma_{\xi}^2}\sum_{m=1}^L a_m^2+\sigma_{\psi}^2+\sigma_{\xi}^2$, passing through the AR model with the coefficients of $\{\frac{a_m\sigma_x^2}{\sigma_x^2+\sigma_{\xi}^2}\}$, $m\in\mathcal{L}$.


According to the results in~\cite{Zamir2008}, the rate distortion function with the source of $y$ is equal to the R-D function with the source of $\nu$. Therefore, we derive the lower bound in the periodic feedback scheme as follows,
\begin{equation}
\begin{aligned}
R(D_x^{})\overset{(a)}{=}&\frac{1}{2L}h(y_{k,1:L})-\frac{1}{2}\log2\pi e \left(D_x-\sigma_P^2\right) +\frac{1}{2}\sum_{l=1}^L\log\theta_l^2\\
\overset{(b)}{=}&\frac{1}{2}\log\frac{\sigma_{\nu}^2}{D_x-\sigma_P^2}+\frac{1}{2}\sum_{l=1}^L\log\theta_l^2,
\end{aligned}
\label{eq: auxiliary RD period solu}
\end{equation}
where $(a)$ follows the results shown in~\cite{Zamir2008}, $(b)$ follows the same derivation in (\ref{eq: RD aperiod solu}).

In the simulations parts of this paper, we will numerically evaluate and compare the theoretical lower bound on the bit rates for different feedback schemes. Next, we will propose practical channel state feedback schemes, which can be implemented by off-the-shelf signal processing devices or modules.

\section{An Practical Channel State Feedback Scheme}
\label{sec: a KF compression}

In the previous section, we calculated the lower bounds on the overhead of the aperiodic and periodic channel state feedback schemes. In this section, we propose practical feedback schemes.

\subsection{High Resolution Quantization in Aperiodic Feedback Scheme} 
\label{subsec: aperiodic compression}


In this subsection, we focus on a practical compression method of CSI in the aperiodic scheme. As already pointed out, the compression method of an aperiodical scheme is essentially a scalar quantization process. The sample-wise quantization has been thoroughly studied. Thus, we pay more attention to evaluating the performance of the practical aperiodical feedback schemes.



Due to its most widely usage in current wireless communication systems, uniform quantizers are considered for aperiodic channel state feedback. The MSE between the channel state $x_{k+1}$ and the uniform quantizer output is taken as the metric to measure the feedback distortion. It is worth noting that the input of the quantizer is the predicted channel state $\hat{x}_{k+1}$. 

Let $\hat{x}_{k+1}^Q$ denote the quantized channel state at the $(k+1)$-th time instance and $e_a$ denote the error between the real channel state and the quantization result. Furthermore, $R_a$ denotes the average number of bits. With the defined variables, we can calculate the MSE as follows, 
\begin{equation}
\begin{aligned}
E[e_a^2]&=E\left[\left(x_{k+1}-\hat{x}_{k+1}^Q\right)^2\right]\\
&=E\left[E\left[\left(x_{k+1}-(\hat{x}_{k+1}+e_a^Q)\right)^2|e_a^Q\right]\right]\\
&=E\left[\begin{matrix}E\left[(x_{k+1}-\hat{x}_{k+1})^2|e_a^Q\right]+(e_a^Q)^2\\
+2E[\left(x_{k+1}-\hat{x}_{k+1})e_a^Q|e_a^Q\right]
\end{matrix}\right]\\
&\overset{(a)}{=}E\left[\sigma_P^2+(e_a^Q)^2+2E\left[(x_{k+1}-\hat{x}_{k+1})e_a^Q|e_a^Q\right]\right]\\
&\overset{(b)}{=}\sigma_P^2+E\left[(e_a^Q)^2\right]\\
&\overset{(c)}{\cong}\sigma_P^2+\frac{1}{12}2^{h(\hat{x})}2^{-2R_a}\\
&=\sigma_P^2+\frac{1}{12}\left(2\pi e(\sigma_x^2+\sigma_P^2)\right)^{\frac{1}{2}}2^{-2R_a}\\
\end{aligned}
\label{eq: aperiodic q mse}
\end{equation}
where $(a)$ follows the result in (\ref{eq: MMSE prediction e}), $(b)$ follows that quantization noise $e_a^Q$ is independent to the prediction error $(x_{k+1}-\hat{x}_{k+1})$ and the mean value of quantization noise is zero, $(c)$ is obtained under the assumption of high-resolution uniform quantization.

In (\ref{eq: aperiodic q mse}), the MSE between the reconstructed channel state at BS $\hat{x}_{k+1}^Q$ and ${x}_{k+1}$ is a function of the number of data bits $R_a$, which can be denoted as a function of MSE through simple mathematical manipulations,
\begin{equation}
R_a=\frac{1}{2}\log\left(\frac{1}{12}\frac{\left(2\pi e(\sigma_x^2+\sigma_P^2)\right)^{\frac{1}{2}}}{E[e_a^2]-\sigma_P^2}\right).
\label{eq: aperiodic R-D uni}
\end{equation}

\subsection{A Practical Channel State Compression Method in Periodic Feedback Scheme} 
\label{subsec: periodic compression}

As shown in Section~\ref{subsec: periodic overhead}, the compressed data of the channel innovation consumes less bits than that for the original channel state. Periodic feedback scheme makes it possible to employ channel innovation based method. In this subsection, we discuss the details of designing a practical method to compress channel innovation under the impact of fading and additive noise. 

\begin{figure}[h!]
\centering
\includegraphics[width=0.8\linewidth] {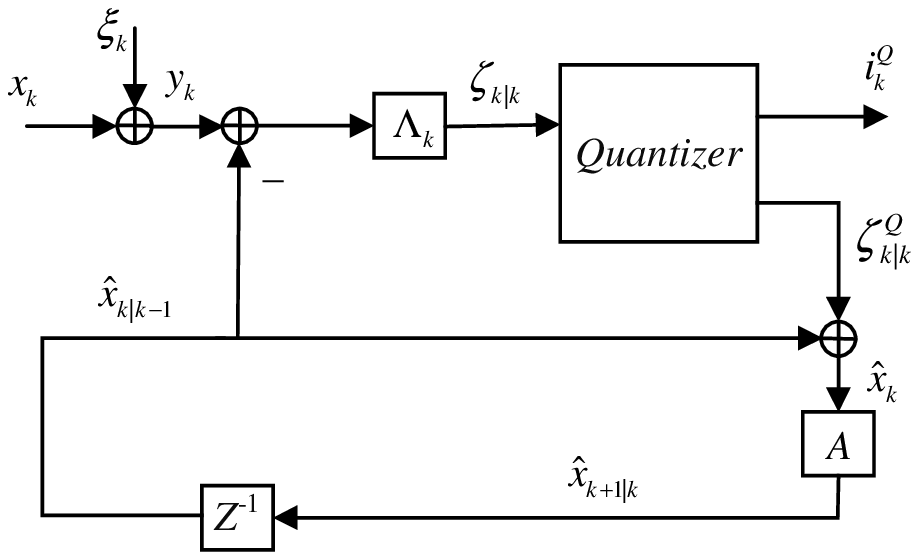}
\setlength{\abovecaptionskip}{3pt plus 3pt minus 2pt}
\caption{Encoder of the proposed additive noise degraded channel innovation compression scheme}
\label{fig:compressor blocks}
\end{figure}
\begin{figure}[h!]
\centering
\includegraphics[width=0.8\linewidth] {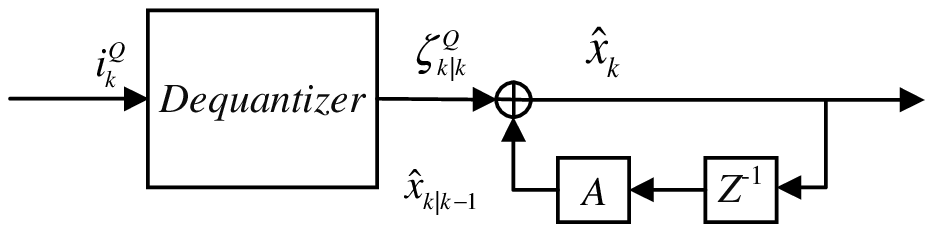}
\setlength{\abovecaptionskip}{3pt plus 3pt minus 2pt}
\caption{Decoder of the proposed compression scheme}
\label{fig:decoder blocks}
\end{figure}

Fig.~\ref{fig:compressor blocks} and Fig.~\ref{fig:decoder blocks} describe the structures of the encoder and decoder of the proposed compression scheme. The optimum configurations and performance analysis will be presented to implement the scheme and evaluate its performance.


As shown in Fig.~\ref{fig:compressor blocks}, the encoder is built on an innovation compression scheme with a feedback loop. The closed loop compression is designed to avoid quantization noise accumulation. In the compression structure, the innovation is denoted by $\zeta$ and its quantization is $\zeta^Q$. The quantized innovation is added to the predicted channel state $\hat{x}_{k|k-1}$ to obtain the estimation of the current channel state $\hat{x}_k$, where $\hat{x}_{k|k-1}$ is the prediction of the channel state based on the previous $L$ channel states $\hat{x}_{k,1:L}$. 



With the reconstructed channel states $\hat{x}_{k+1,1:L}=\{\hat{x}_{k-L+1},\hat{x}_{k-L+2},\cdots,\hat{x}_{k}\}$, the channel state at the $(k+1)$-th time instance is predicted as follows,
\begin{equation}
\hat{x}^{}_{k+1|k}=\sum_{m=1}^L a_m \hat{x}^{}_{k-m+1}.
\label{eq: self predict}
\end{equation}

Passing $\hat{x}^{}_{k+1|k}$ through a time delay unit, we have the output $\hat{x}^{}_{k|k-1}$. $\hat{x}^{}_{k|k-1}$ is the estimation of $x_k$ based on $\hat{x}_{k,1:L}=\{\hat{x}_{k-L},\hat{x}_{k-L+1},\cdots,\hat{x}_{k-1}\}$. 


MSE of the prediction, $E[e^2_{k|k-1}]$, where $e_{k|k-1}=x_k-\hat{x}_{k|k-1}$, can be calculated as follows,
\begin{equation}
E[e^2_{k|k-1}]=\sum_{m=1}^L a_m^2 E[e^2_{k-m}]+\sigma_{\psi}^2+\sum_{m=1}^L a_m^2 E[{e^Q}^2_{k-m}].
\label{eq: P predict}
\end{equation}

\begin{proof}
See Appendix~\ref{app_proof_mse_predict}.
\end{proof}


$\hat{x}^{}_{k|k-1}$ is the initial estimation of $x_k$. The accuracy in the prediction of $x_k$ is affected by two factors: the estimation error and quantization noise. Therefore, we need to update $\hat{x}^{}_{k|k-1}$ in the closed loop to avoid the accumulation of noise induced by these two factors:
\begin{equation}
\hat{x}_{k}=\hat{x}^{}_{k|k-1}+\zeta_k^Q,
\label{eq: chan state update}
\end{equation}
where $\zeta_k^Q$ is the quantization output of $\zeta_k$.

Next, we discuss how to obtain $\zeta_k^Q$. The MMSE of the prediction errors with respect to $\zeta_k^Q$ is defined as follows
\begin{equation}
\min_{\zeta_k^Q}E\left[(x_k-\hat{x}^{}_{k|k-1})^2\right].
\label{eq: updating rule}
\end{equation}


Since a fading channel is essentially a time evolution random process, an iterative solution to (\ref{eq: updating rule}) can be obtained to dynamically approach the optimum value as follows, 
\begin{equation}
\zeta_k^Q=E[e^2_{k|k-1}]\left(E[e^2_{k|k-1}]+\sigma^2_{\xi}\right)^{-1}\left(y_k-\hat{x}^{}_{k|k-1}\right)，
\label{eq: opt zeta}
\end{equation}
where $e_{k|k-1}=x_k-\hat{x}^{}_{k|k-1}$.

\begin{proof}
See Appendix~\ref{app_proof_kalman_gain}.
\end{proof}

Via the updating method shown in (\ref{eq: chan state update}), MMSE is asymptotically achieved in an iterative manner. The two mean square errors, $E[e_k^2]$ and $E[e_{k|k-1}^2]$, can thus satisfy the following condition. 
\begin{equation}
E\left[e^2_k\right]=\left(1-\lambda_k\right)E\left[e^2_{k|k-1}\right],
\label{eq: kalman gain}
\end{equation}
where $\lambda_k=E[e^2_{k|k-1}]\left(E[e^2_{k|k-1}]+\sigma^2_{\xi}\right)^{-1}$


With the optimum innovation $\zeta_k^Q$, the transmitter can then use the quantization codebook to determine the corresponding index and map it to a symbol for transmission. 

At the decoder side, the symbol is first demapped to obtain the codeword index, which is then used to reconstruct the quantized innovation $\zeta_k^Q$. With the reconstructed $\zeta_k^Q$, we can calculate the channel state using the structure presented in Fig.~\ref{fig:decoder blocks}. To distinguish from the estimated channel state at the encoder side $\hat{x}_k$, the reconstructed channel state at the decoder side is denoted by $\hat{x}_k^*$. As shown in Fig.~\ref{fig:decoder blocks}, the reconstructed $\zeta_k^Q$ is added to $x_k$ which is the prediction based on the previous $L$ channel states:
\begin{equation}
\hat{x}^*_k=\sum_{m=1}^L a_m \hat{x}^*_{k-m}+\zeta^Q_k.
\label{eq: decoder reconstruct}
\end{equation}

After introducing the the encoder and decoder of the proposed channel state feedback scheme, we will investigate its performance in the next subsection. 

\subsection{Performance of the Proposed Feedback Scheme}
\label{subsec: theoretic bound}

The same performance metric used in the previous part of this paper: bit-distortion curve is taken to evaluate the performance of the proposed feedback scheme. We first prove that the MSE converges in the long term, and then calculate this long-term MSE, which is defined as follows,


\begin{equation}
\varsigma=\lim_{k\rightarrow\infty} E\left[\left(x_k-\hat{x}_k\right)^2\right]
\label{eq: stable mse def}.
\end{equation}

Theorem~\ref{theorem: diff eq mse} shows that the long-term MSE $\varsigma$ can be expressed in an iterative structure. 

\begin{Theorem}
\label{theorem: diff eq mse}
The stable MSE $\varsigma$ is the solution to the differential equation as follows,
\begin{equation}
\varsigma-\left((1-\lambda)^2\sum_{m=1}^L a^2_m \varsigma^{(m)}\right)=(1-\lambda_k)^2\sigma_{\psi}^2+\lambda_k^2\sigma_{\xi}^2+\sigma_{Q}^2.
\label{eq: diff varsigma}
\end{equation}
where $\varsigma^{(m)}$ denotes the $m$-th order derivative of $\varsigma$.
\end{Theorem}

\begin{proof}
See Appendix~\ref{app_proof_mse_diff}.
\end{proof}

Let
\begin{equation}
\sum_{m=1}^L f_m e^{c_m nT}
\label{eq: general solu}
\end{equation}
and 
\begin{equation}
\sum_{m=0}^L g_m (nT)^m
\label{eq: particular solu}
\end{equation}
denote the general and specific solutions to the differential equation (\ref{eq: diff eq}). Combining (\ref{eq: general solu}) and (\ref{eq: particular solu}), the solution to (\ref{eq: diff eq}) is written as follows,
\begin{equation}
\varsigma^*=\sum_{m=1}^L f_m e^{c_m nT}+\sum_{m=0}^L g_m (nT)^m.
\label{eq: diff eq}
\end{equation}

However, it is extremely difficult to obtain the explicit forms of $\{f_m\}$, $\{c_m\}$, and $\{g_m\}$, $m\in\mathcal{L}$. In order to gain insight on the long-term behaviour of MSE $\varsigma$, we tackle this problem by introducing the approximation of the channel state process in an order one autoregressive model (AR(1)). Let $\tilde{x}$ denote the approximation of $x$,
\begin{equation}
\tilde{x}_k=\beta\tilde{x}_{k-1}+\iota_k.
\label{eq: tilde x def}
\end{equation}
where the coefficient $\beta$ is determined under the rule of MMSE, that is,
\begin{equation}
\beta=\min_{\beta}E\left[\left(x_k-\beta\tilde{x}_{k-1}\right)^2\right],
\label{eq: mmse approx}
\end{equation}
and $\iota$ is a zero mean Gaussian random process with the variance of $\sigma_{\iota}^2$ to be determined.

From (\ref{eq: mmse approx}), we obtain the value of $\beta$ as follows,
\begin{equation}
\beta=\frac{\kappa_x(1)}{\kappa_x(0)}.
\label{eq: beta cal}
\end{equation}

With the calculated $\beta$, the variance $\sigma_{\iota}^2$ is correspondingly determined,
\begin{equation}
\sigma_{\iota}^2=\kappa_x(0)-\frac{\kappa_x^2(1)}{\kappa_x(0)}.
\label{eq: sigma iota cal}
\end{equation}

Afterwards, we calculate the MSE of the approximated solution. Following a similar method as (\ref{eq: diff varsigma}), we obtain a first order differential equation as follows,
\begin{equation}
\tilde{\varsigma}_k-(1-\lambda_k)^2 \beta^2 \tilde{\varsigma}_{k-1}^{}=(1-\lambda_k)^2\sigma_{\iota}^2+\lambda_k^2\sigma_{\xi}^2+\sigma_{Q}^2.
\label{eq: diff varsigma approx}
\end{equation}

From (\ref{eq: diff varsigma approx}), the long-term value of $\tilde{\varsigma}$ can be calculated as follows,
\begin{equation}
\begin{aligned}
\tilde{\varsigma}_{\infty}=&\lim_{K\rightarrow\infty}\left(\begin{matrix}
\left((1-\lambda_k)^2 \beta^2\right)^K\tilde{\varsigma}_0\\
+\left(\begin{matrix}\left((1-\lambda_k)^2\sigma_{\iota}^2+\lambda_k^2\sigma_{\xi}^2+\sigma_{Q}^2\right)\\
\times\frac{1-\left((1-\lambda_k)^2 \beta^2\right)^K}{1-(1-\lambda_k)^2 \beta^2}
\end{matrix}\right)
\end{matrix}\right)
\\
\overset{(a)}{=}&\frac{(1-\lambda_k)^2\sigma_{\iota}^2+\lambda_k^2\sigma_{\xi}^2+\sigma_{Q}^2}{1-(1-\lambda_k)^2 \beta^2},\\
\label{eq: varsigma inf cal}
\end{aligned}
\end{equation}
where $(a)$ follows $0<(1-\lambda_k)<1$ and $0<\beta<1$.

After calculating the long-term MSE, we investigate the quantization bits in our analysis to obtain the bit versus MSE curve. In the proposed feedback scheme, the quantization objective is essentially equal to the prediction error $e_P$, i.e., $e_P=x_k-E[x_k|y_k-1]$. Thus, the statistic feature of $e_P$ is a function of the quantization bit number. However, we can hardly calculate the explicit distribution function of $e_P$. As an extreme case, when $e_P$ follows a Gaussian distribution, the quantization of $e_P$ requires the largest number of bits in order to meet the same level of errors. We can assume the quantization objective follows the zero mean Gaussian distribution.

To estimate the bit number, we calculate the variance of $e_P$ under the Gaussian distribution assumption. The calculation of $E[e_P^2]$ is derived as follows,
\begin{equation}
\begin{aligned}
E[\tilde{e}_P^2]=&\kappa_{\tilde{x}}(0)-\frac{\kappa_{\tilde{x}y}^2(1)}{\kappa_{\tilde{x}y}(0)+\sigma_{\xi}^2}\\
=&\kappa_{\tilde{x}}(0)-\frac{\kappa_{\tilde{x}}^2(1)}{\kappa_{\tilde{x}}(0)+\sigma_{\xi}^2}.
\end{aligned}
\label{eq: app min predict mse}
\end{equation}
where 
\begin{equation}
\begin{aligned}
\kappa_{\tilde{x}}(0)=\int_{-\pi}^{\pi}\left(\kappa_x(0)-\frac{\kappa_x^2(1)}{\kappa_x(0)}\right)\frac{1}{|1-\beta e^{j\omega}|^2}d\omega.
\end{aligned}
\label{eq: kappa tilde x 0}
\end{equation}
\begin{equation}
\begin{aligned}
\kappa_{\tilde{x}}(1)&=\beta\int_{-\pi}^{\pi}\left(\kappa_x(0)-\frac{\kappa_x^2(1)}{\kappa_x(0)}\right)\frac{1}{|1-\beta e^{j\omega}|^2}d\omega\\
&=\int_{-\pi}^{\pi}\left(\kappa_x(1)-\frac{\kappa_x^3(1)}{\kappa^2_x(0)}\right)\frac{1}{|1-\beta e^{j\omega}|^2}d\omega.\\
\end{aligned}
\label{eq: kappa tilde x 1}
\end{equation}

Next, we aim to work out the relationship between the number of quantization bits and the quantization error. In uniform quantization, the mean square of the quantization error is denoted by $\sigma_{Q^u}^2$. Let $R_u$ denote the average number of bits for quantization. Under the high resolution quantization assumption, $R_u$ and $\sigma_{Q^u}^2$ satisfy the following equation,
\begin{equation}
\begin{aligned}
\sigma_{Q^u}^2&\overset{(a)}{\cong}\frac{1}{12}2^{h(e_P)}2^{-2R_u}\\
&=\frac{1}{12}2^{\frac{1}{2}\log 2\pi eE[e_P^2]}2^{-2R_u}\\
\end{aligned}
\label{eq: high res mse},
\end{equation}

Substituting the corresponding parts of (\ref{eq: high res RD}) with (\ref{eq: varsigma inf cal}) and (\ref{eq: app min predict mse}), we have
\begin{equation}
\begin{aligned}
R_{u}&=\frac{1}{2}\log\frac{1}{12}+\frac{1}{2}\log\sqrt{2\pi e\kappa_{\tilde{x}}(0)-\frac{\kappa_{\tilde{x}}^2(1)}{\kappa_{\tilde{x}}(0)+\sigma_{\xi}^2}}\\
&-\frac{1}{2}\log\left(\left({1-(1-\lambda_k)^2 \beta^2}\right)\tilde{\varsigma}_{\infty}-(1-\lambda_k)^2\sigma_{\iota}^2+\lambda_k^2\sigma_{\xi}^2\right).
\end{aligned}
\label{eq: high res RD}
\end{equation}

From (\ref{eq: high res RD}), we can easily know the average number of bits $R_u$ at a given prediction MSE $\tilde{\varsigma}_{\infty}$, or the inverse relation between $\tilde{\varsigma}_{\infty}$ and $R_u$.

\section{Numerical Simulations}
\label{sec: simulations}


In this section, we first numerically study the theoretical lower bounds of different feedback schemes. Afterwards, extensive numerical simulations have been performed to evaluate the proposed practical feedback method. In the simulation, we consider a more practical fading channel, which is the output of an AR(4) model with a white Gaussian random variable. The same assumption that the channel $x$ cannot not be directly observable at the receiver due to the AWGN $\xi$ is introduced. Therefore, four previous noisy channel observations are exploited to predict the one-step ahead channel state. 


The traditional ZH method is also simulated in order to provide performance comparison. Furthermore, the theoretical lower bounds for both aperiodical and periodical feedback schemes are studied. As discussed in the previous sections, in the aperiodical scheme, the original channel states are quantized for feedback, while in the periodical one, channel innovations are quantized and sent back to the base station.

\begin{figure}[h!]
\centering
\includegraphics[width=1\linewidth] {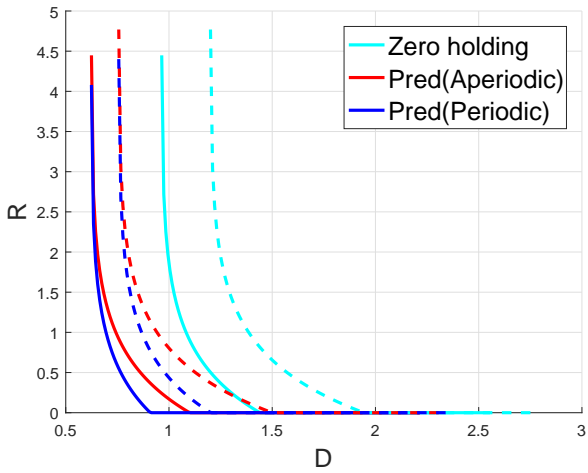}
\setlength{\abovecaptionskip}{3pt plus 3pt minus 2pt}
\caption{Bit rate versus channel distortion under varied input variances to the channel AR model.}
\label{fig: RD vary psi}
\end{figure}

First, we calculate the bit rate $R$ versus channel distortion $D$ for all the three scemes: \textit{zero-holding}, \textit{prediction+aperiodical}, and \textit{prediction+periodical}. The R-D curves are generated by varying the input variance of the AR model. Fig.~\ref{fig: RD vary psi} shows the results. 
 
In Fig.~\ref{fig: RD vary psi}, the solid lines are the results with $\sigma_{\psi}^2=1$. Remember that $\psi$ is the white Gaussian variable which is the input to the AR model to generate the fading channel. The dashed lines correspond to $\sigma_{\psi}^2=2$. 

From Fig.~\ref{fig: RD vary psi}, we can observe that the number of bits deceases with the increase of channel state reconstruction error. Furthermore, there is a bound on the distortion in curve, that is, the reconstruction error can not be smaller than a threshold value even if the number of bits is infinitely large. This bound is determined by the finite channel prediction accuracy which is irrelevant to the channel state feedback bits. 

The ZH feedback scheme requires more bits at the same distortion level compared with the proposed schemes. This disadvantage is caused by the lower prediction accuracy since it simply takes the channel state at the current time instance for the next time stance. Furthermore, among \textit{aperiodical + prediction} and \textit{periodical + prediction}, the latter requires less bits since only innovations between consecutive channel states are transmitted and the innovations usually have smaller entropy than the original channel samples. It is worth noting that the two curves overlap with each other when the number of bits becomes sufficiently large. In the extreme case of infinite bits quantization, no channel information loss will be caused by the quantization, therefore, the distortion is purely determined by the prediction error. 

Furthermore, from Fig.~\ref{fig: RD vary psi}, all curves shift right when the channel variance increases. The reason is that increased variance raises the channel entropy, thus more bits are needed at the same distortion level.


\begin{figure}[h!]
\centering
\includegraphics[width=1\linewidth] {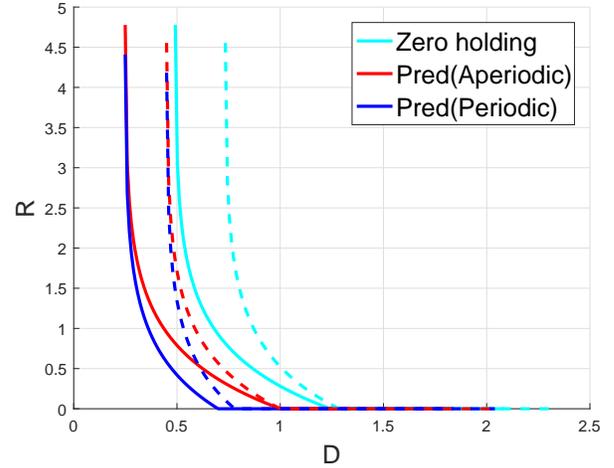}
\setlength{\abovecaptionskip}{3pt plus 3pt minus 2pt}
\caption{Bit rate versus distortion under varied additive Gaussian noise}
\label{fig: RD vary xi}
\end{figure}

Fig.~\ref{fig: RD vary xi} presents the R-D curves  where the variance of the channel model is fixed, while the strength of the additive noise varies. Generally, the number of bits decreases with the increase of noise strength. The ZH scheme generates the largest distortion at the same noise level, compared with the other two schemes. All curves have bounds on distortion when the number of bits increases to the infinity. The \textit{prediction + aperiodical} and \textit{prediction+periodical} schemes have the same bounds. 


The major difference between Fig. \ref{fig: RD vary xi} and Fig.~\ref{fig: RD vary psi} lies in the two curves of the \textit{prediction + aperiodical} scheme. For different level of additive Gaussian noise, the bit rate approaches to zero at the same value of distortion. When no bit is used to transmit the channel states, the distortion is purely determined by the variance of the channel. Since the two curves with different levels of noise have the same channel variance, the distortion at zero bit rate is the same. However, the variance of channel innovations is related to the additive noise. Therefore, the R-D curves of the \textit{prediction+aperiodical} scheme have different zero-bits distortions if the additive noise is not the same.

\begin{figure}[h!]
\centering
\includegraphics[width=1\linewidth] {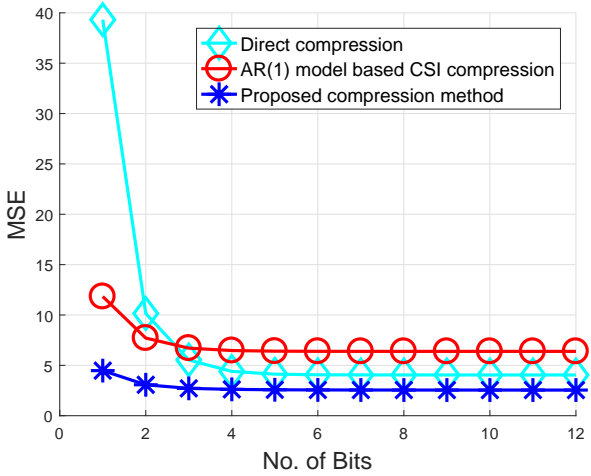}
\setlength{\abovecaptionskip}{3pt plus 3pt minus 2pt}
\caption{Number of bits versus channel reconstruction MSE in the proposed CSI feedback schemes}
\label{fig: practical RD}
\end{figure}

In order to investigate the performance of the proposed CSI feedback scheme, we calculate the curves of the channel reconstruction MSE versus the number of feedback bits. The experiment results are plotted in Fig.~\ref{fig: practical RD} for the following three schemes: 1) the channel state is directly quantized and sent back to the base station. This method is widely applied in current systems; 2) the channel is modeled as an AR(1) model without considering additive noise. Under such model, the innovation between two consecutive channel states is quantized and sent back to the base station; 3) a more practical scheme with a high-order AR model under AWGN. The compression method proposed in Section~\ref{subsec: periodic compression} is used for transmitting CSI. 

Fig.~\ref{fig: practical RD} shows that, the direct quantization scheme generates the largest channel reconstruction MSE when the number of bits is small. If the feedback bits increases, the reconstruction MSE decreases and reaches a stable level that is lower than the AR(1) model. The direct quantization scheme's MSE is determined by the additive noise imposed on the fading channel. The MSE of the AR(1) model is always larger than the proposed method, which is caused by the following two factors: first, the AR(1) model is not accurate and introduces some error when expressing a fading channel; second, the additive Gaussian noise is not considered and adds to the quantization error. 

\section{Conclusions}
\label{sec: conclusion}

In this paper, we investigate the overhead introduced by CSI feedback both theoretically and practically, under the periodical and aperiodical CSI feedback schemes that are employed by the LTE standard and discussed in designing future 5G systems. Within these two schemes, we calculate the theoretical lower bounds on the bits for conveying channel states. We observe that the bound of the periodical feedback scheme is lower than the aperiodical scheme, because the channel innovation feedback removes the redundancy so that less bits are needed. We also consider a more practical channel model than the AR(1) model widely accepted in current research on CSI feedback, as well as the additive noise that deteriorates channel states, and propose a practical CSI feedback scheme, which periodically sends the channel state innovation estimated from the noisy observation of channel states in a highly efficient and effective manner. Simulations show that the proposed CSI feedback scheme outperforms those in literature, and suggest its potentially significant contribution to the future generation of wireless communication systems.

\vspace{-3mm}
\appendix
\subsection{Proof of equation (\ref{eq: P predict})}
\label{app_proof_mse_predict}

\begin{proof}
The error in the prediction of $x_k$ based on $\bold{x}_{k-L;L}$ where $\bold{x}_{k-L;L}=[x_{k-L}, x_{k-L+1}, \cdots, x_{k-1}]$ is defined as follows,
\begin{equation}
\begin{aligned}
e_{k|k-1}&=x_k-\hat{x}^{\prime}_{k|k-1}\\
&=\bold{a}^T\bold{x}_{k-L;L}+ \psi_k -\bold{a}^T\hat{\bold{x}}_{k-L;L}-\bold{a}^T\bold{e}^Q_{k-L;L}\\
&=\bold{a}^T \bold{e}_{k-L;L} + \psi_k -\bold{a}^T\bold{e}^Q_{k-L;L},\\
\end{aligned}
\label{eq: e k k-1 def}
\end{equation}
where 
\begin{equation}
\begin{aligned}
&\bold{a}=[a_1,a_2,\cdots,a_L]^T\\
&\hat{\bold{x}^T}_{k-L;L}=[\hat{x}_{k-L}, \hat{x}_{k-L+1}, \cdots, \hat{x}_{k-1}]^T\\
&\bold{e}^Q_{k-L;L}=[e^Q_{k-L}, e^Q_{k-L+1}, \cdots, e^Q_{k-1}]^T\\
&\bold{e}_{k-L;L}=[e_{k-L}, e_{k-L+1}, \cdots, e_{k-1}]^T.
\end{aligned}
\end{equation}

Therefore, the mean square error in the prediction of $x_k$ based on the $L$ past observations is calculated as follows,
\begin{equation}
\begin{aligned}
E[e^2_{k|k-1}]&=E\left[e_{k|k-1}e^H_{k|k-1}\right]\\
&=E\left[\left(\bold{a}^T \bold{e}_{k-L;L} + \psi_k -\bold{a}^T\bold{e}^Q_{k-L;L}\right)\right.\\
&~~~\cdot\left.\left(\bold{a}^T \bold{e}_{k-L;L} + \psi_k -\bold{a}^T\bold{e}^Q_{k-L;L}\right)^H\right]\\
&\overset{(a)}{=}E\left[\bold{a}^T\bold{e}_{k-L;L}\bold{e}^H_{k-L;L}\bold{a}\right]+E\left[\psi_k^2\right]\\
&~~~+E\left[\bold{a}^T\bold{e}^Q_{k-L;L}{\bold{e}^Q}H_{k-L;L}\bold{a}\right]\\
&\overset{(b)}{=}E\left[\bold{a}^T diag\{e^2_{k-L},e^2_{k-L+1},\cdots, e^2_{k-1},\}\bold{a}\right]+E\left[\psi_k^2\right]\\
&+E\left[\bold{a}^T diag\{{e^Q}^2_{k-L},{e^Q}^2_{k-L+1},\cdots, {e^Q}^2_{k-1},\}\bold{a}\right]\\
&=\sum_{m=1}^L b_m^2 e^2_{k-m} ++\sigma_{\psi}^2+\sum_{m=1}^L b_m^2 {e^Q}^2_{k-m}
\end{aligned}
\label{eq: mse k k-1}
\end{equation}
where $(a)$ follows that the estimation error $e$　is independent to both the additive noise $\psi$ and the quantization noise $e^Q$, and the quantization noise $e^Q$ is independent to $\psi$; $(b)$ follows that
\begin{equation}
E[e_{k_1} e_{k_2}]=\left\{\begin{matrix}e^2_{k_1}, & {k_1}={k_2}\\
0, & {k_1}\neq{k_2}\end{matrix}\right.
\label{eq: e m e n}
\nonumber
\end{equation}
and 
\begin{equation}
E[e^Q_{k_1} e^Q_{k_2}]=\left\{\begin{matrix}\sigma^2_Q, & {k_1}={k_2}\\
0, & {k_1}\neq{k_2}\end{matrix}\right..
\label{eq: e m e n}
\nonumber
\end{equation}
Thus, the proof is completed.
\end{proof}

\subsection{Proof of equation (\ref{eq: chan state update})}
\label{app_proof_kalman_gain}

\begin{proof}

In the iterative calculation for the optimum updating, the noisy channel state $y_k$ is the only variable which can be observed. The noisy observation is taken to assist the optimum updating. Thus, we first write the difference between the predicted noisy observation and the real observation $y_k$, which is also called as the innovation, as follows,
\begin{equation}
\mu_{k}=y_k-\hat{x}^{}_{k|k-1}.
\label{eq: innovation cal}
\end{equation}

The calculated $\mu_{k}$ is essentially observed value which is not optimized. To achieve the updating in rule of MMSE, an modification on $\mu_{k}$ is needed. We consider the multiplicative factor to modify $\mu_{k}$ such that the MMSE updating can be achieved. In the second step, we thus calculate the product between the gain $\lambda_k$ and $\mu_k$ as follows,
\begin{equation}
\zeta_{k}=\lambda_k\mu_k,
\label{eq: weighted information}
\end{equation}
where $\lambda_k$ is denotes the gain.

The quantized $\zeta_k$, denoted by $\zeta_k^Q$, is used in the MMSE updating. Thus, the quantized innovation $\zeta_k^Q$ is used in the following calculation of MSE. In the quantization of $\zeta_k$, we consider uniform quantization due to its most widely use. The mean square error of $x_k$ with respect to $\hat{x}_k$ is calculated as follows,
\begin{equation}
\begin{aligned}
E[{e}^2_{k}]&=E\left[e_k e^H_k\right]\\
&=E\left[\left(x_k-\hat{x}_k\right) \left(x_k-\hat{x}_k\right)^H_k\right]\\
&\overset{(a)}{=}E\left[\left(x_k+\xi_k- \hat{x}^{\prime}_{k|k-1}-\lambda_k(y_k-\hat{x}^{}_{k|k-1})-e_k^Q\right)\right.\\
&~~\left.\cdot \left(x_k+\xi_k-\hat{x}^{}_{k|k-1} -\lambda_k(x_k+\xi_k-\hat{x}^{}_{k|k-1})-e_k^Q\right)^H\right]\\
&\overset{(b)}{=}E\left[\left(1-\lambda_k\right)\left(x_k-\hat{x}^{}_{k|k-1}\right)\left(x_k-\hat{x}^{}_{k|k-1}\right)^H \left(1-\lambda_k\right)^H\right.\\
&~~\left.+\lambda_k\xi_k\xi_k^H\lambda_k^H + e_k^Q (e_k^Q)^H\right]\\
&=E\left[\left(1-\lambda_k\right){e}^2_{k|k-1} \left(1-\lambda_k\right)^H+\lambda_k \sigma_{\xi}^2\lambda_k^H + \sigma_{Q}^2\right],
\end{aligned}
\label{eq: mse e2 k}
\end{equation}
where $(a)$ follows (\ref{eq: 1 observation}), (\ref{eq: innovation cal}), (\ref{eq: weighted information}), and (\ref{eq: chan state update}); $\zeta_k^Q$ is the error in quantizing $\zeta_k$. $(b)$ follows the fact that the additive noise $\xi$ is independent to all of three terms: the channel state $x$, the prediction of $x$ based on their observations in past time instances, and the quantization noise $e^Q$; furthermore, $(b)$ is obtained according to the weak assumption that the quantization noise is independent to the channel state $x$ and its linear estimation.

Next, we calculate the derivative of $E[{e}^2_{k}]$ with respect to $\lambda_k$ and calculate the $\lambda_k$ generating the zero derivative,
\begin{equation}
\frac{d E[e^2_k]}{d \lambda_k}=0.
\label{eq: 0 derivative def}
\end{equation}


After solving (\ref{eq: 0 derivative def}), we have the optimum gain shown as follows, 
\begin{equation}
\lambda_k^*=E[e^2_{k|k-1}]\left(E[e^2_{k|k-1}]+\sigma^2_{\xi}\right)^{-1}，
\label{eq: solved lambda}
\end{equation}
where $E[e^2_{k|k-1}]$ denotes the mean square error matrix calculated in the previous round of compression.

After combining (\ref{eq: innovation cal}), (\ref{eq: weighted information}), and (\ref{eq: solved lambda}), the proof is completed.
\end{proof}
 
\subsection{Proof of Theorem~\ref{theorem: diff eq mse}}
\label{app_proof_mse_diff}

Before calculating $\varsigma_k$, we rewrite $\hat{x}_k$ into an iterative structure. Combining (\ref{eq: chan state update}), (\ref{eq: kalman gain}), (\ref{eq: weighted information}) and (\ref{eq: innovation cal}), we have the reconstruction of channel state at $k$-th time instance,
\begin{equation}
\begin{aligned}
\hat{x}_k=&\hat{x}_{k|k-1}^{}+\lambda_k\left(y_k-\hat{x}_{k|k-1}^{}\right)+e^Q_k\\
=&\sum_{m=1}^L a_m\hat{x}_{k-m}\\
+&\lambda_k\left(\sum_{m=1}^L a_m \left(x_{k-m}-\hat{x}_{k-m}\right)+\psi_k+\xi_k\right)+e^Q_k,\\
\end{aligned}
\label{eq: hat xk cal1}
\end{equation}
where $e_k^Q$ denotes the quantization noise.

Afterwards, we calculate the long-term reconstruction MSE at $k$-th time instance and write it into an iterative structure. The derivations are as follow
\begin{equation}
\begin{aligned}
\varsigma_k=&E\left[\left(x_k-\hat{x}_k\right)^2\right]\\
=&E\left[\left(\sum_{m=1}^L a_m{x}_{k-m}+\psi_k-\sum_{m=1}^L a_m\hat{x}_{k-m}\right.\right.\\
-&\left.\left.\lambda_k\left(\sum_{m=1}^L a_m \left(x_{k-m}-\hat{x}_{k-m}\right)+\psi_k+\xi_k\right)-e^Q_k\right)^2\right]\\
\overset{(a)}{=}&(1-\lambda_k)^2E\left[\left(\sum_{m=1}^L a_m({x}_{k-m}-\hat{x}_{k-m})\right)^2\right]\\
+&(1-\lambda_k)^2\sigma_{\psi}^2+\lambda_k^2\sigma_{\xi}^2+\sigma_{Q}^2\\
\overset{(b)}{=}&(1-\lambda_k)^2E\left[\left(\sum_{m=1}^L a_m^2({x}_{k-m}-\hat{x}_{k-m})^2\right)\right]\\
+&(1-\lambda_k)^2\sigma_{\psi}^2+\lambda_k^2\sigma_{\xi}^2+\sigma_{Q}^2\\
=&(1-\lambda_k)^2E\left[\left(\sum_{m=1}^L a_m^2\varsigma_{k-m}\right)\right]\\
+&(1-\lambda_k)^2\sigma_{\psi}^2+\lambda_k^2\sigma_{\xi}^2+\sigma_{Q}^2
\end{aligned}
\label{eq: stable k-th mse cal},
\end{equation}
where $(a)$ follows the facts, $\psi$, $\xi$, and $e^Q$ are independent to each other; $\psi_{k_1}$ is uncorrelated with the channel states $x_{k_2}$ and noisy channel observations $y_{k_2}$ before the $k_1$-th time instance, $k_1>k_2$; Similarly, $\xi_{k_1}$ is also uncorrelated with the channel states $x_{k_2}$ and noisy channel observations $y_{k_2}$ for $k_1>k_2$; $(b)$ follows the equation below
\begin{equation}
\begin{aligned}
&E\left[\left(\sum_{m=1}^L a_m({x}_{k-m}-\hat{x}_{k-m})\right)^2\right]\\
=&E\left[\left(\sum_{m=1}^L a_m^2({x}_{k-m}-\hat{x}_{k-m})^2\right)\right].
\end{aligned}
\label{eq: lemma diag}
\end{equation}
which follows the result in Lemma~\ref{lemma: zero cross term}.

From (\ref{eq: stable k-th mse cal}), we build a differential equal shown in Theorem~\ref{theorem: diff eq mse}.

Thus, the proof is completed.

\begin{Lemma}
\label{lemma: zero cross term}
Cross terms in the mean square error between linear combinations are approximately equal to zero
\begin{equation}
\begin{aligned}
E\left[\left({x}_{k_1}-\hat{x}_{k_1}\right)\left({x}_{k_2}-\hat{x}_{k_2}\right)\right]=0,
\end{aligned}
\label{eq: zero cross}
\end{equation}
where $k_1\neq k_2$.
\end{Lemma}

\begin{proof}
According to the derivation of (\ref{eq: 0 derivative def}), the $\hat{x}$ is a minimum mean square error estimation of $x$ based on its noisy observation $y$. According to the orthogonal principle, the estimation error $(x-\hat{x})$ is orthogonal to $y$. Furthermore, $\hat{x}$ is a linear combination of $y$. Thus, $(x-\hat{x})$ is also orthogonal to $\hat{x}$, that is,
\begin{equation}
E[(x-\hat{x})\hat{x}]=0.
\label{eq: orth idea}
\end{equation}

We can straightforwardly prove the mean of the estimation error is zero. Furthermore, via a numeric method, we can show the estimation error at a time instance, say $k_1$, is independent to the channel state at a different time instance, say $k_2$, $k_1\neq k_2$. Therefore, we have
\begin{equation}
E[(x_{k_1}-\hat{x}_{k_1})\hat{x}_{k_2}]=E[(x_{k_1}-\hat{x}_{k_1})]E[\hat{x}_{k_2}]=0.
\label{eq: uncor}
\end{equation}

According to the results in (\ref{eq: orth idea}) and (\ref{eq: uncor}), the cross terms in $E\left[\left(\sum_{m=1}^L a_m({x}_{k-m}-\hat{x}_{k-m})\right)^2\right]$ are equal to zero. Thus, the proof of the Lemma is completed.
\end{proof}

\linespread{1.27}
\bibliography{chann_emu_InfoCom2013}

\begin{thebibliography}{10}
\providecommand{\url}[1]{#1}
\csname url@samestyle\endcsname
\providecommand{\newblock}{\relax}
\providecommand{\bibinfo}[2]{#2}
\providecommand{\BIBentrySTDinterwordspacing}{\spaceskip=0pt\relax}
\providecommand{\BIBentryALTinterwordstretchfactor}{4}
\providecommand{\BIBentryALTinterwordspacing}{\spaceskip=\fontdimen2\font plus
\BIBentryALTinterwordstretchfactor\fontdimen3\font minus
  \fontdimen4\font\relax}
\providecommand{\BIBforeignlanguage}[2]{{%
\expandafter\ifx\csname l@#1\endcsname\relax
\typeout{** WARNING: IEEEtran.bst: No hyphenation pattern has been}%
\typeout{** loaded for the language `#1'. Using the pattern for}%
\typeout{** the default language instead.}%
\else
\language=\csname l@#1\endcsname
\fi
#2}}
\providecommand{\BIBdecl}{\relax}
\BIBdecl

\bibitem{Huang2014}
P.~Huang and D.~Rajan, ``Estimation of centralized spectrum sensing overhead
  for cognitive radio networks,'' in \emph{2014 IEEE 25th Annual International
  Symposium on Personal, Indoor, and Mobile Radio Communication (PIMRC)}, Sept
  2014, pp. 659--663.

\bibitem{Huang2011}
P.~Huang and Y.~Pi, ``Wireless internet assisting satellite position in urban
  environments,'' in \emph{2011 6th International ICST Conference on
  Communications and Networking in China (CHINACOM)}, Aug 2011, pp. 262--267.

\bibitem{Huang2011a}
------, ``Urban environment solutions to gps signal near-far effect,''
  \emph{IEEE Aerospace And Electronic Systems Magazine}, vol.~26, no.~5, pp.
  18--27, 2011.

\bibitem{Huang2013b}
P.~Huang, Y.~Pi, and I.~Progri, ``Gps signal detection under multiplicative and
  additive noise,'' \emph{Journal of Navigation}, vol.~66, no.~04, pp.
  479--500, 2013.

\bibitem{Lakshmana2016}
T.~R. Lakshmana, A.~Tölli, R.~Devassy, and T.~Svensson, ``Precoder design with
  incomplete feedback for joint transmission,'' \emph{IEEE Transactions on
  Wireless Communications}, vol.~15, no.~3, pp. 1923--1936, March 2016.

\bibitem{Anand2017}
K.~Anand, E.~Gunawan, and Y.~L. Guan, ``Precoder designs for the relay-aided x
  channel without source csi,'' \emph{IEEE Transactions on Signal Processing},
  vol.~65, no.~1, pp. 41--55, Jan 2017.

\bibitem{Wang2016}
H.~Wang, R.~Song, and S.~H. Leung, ``Throughput analysis of interference
  alignment for a general centralized limited feedback model,'' \emph{IEEE
  Transactions on Vehicular Technology}, vol.~65, no.~10, pp. 8775--8781, Oct
  2016.

\bibitem{Rezaee2016}
M.~Rezaee and M.~Guillaud, ``Interference alignment with quantized grassmannian
  feedback in the k-user constant mimo interference channel,'' \emph{IEEE
  Transactions on Wireless Communications}, vol.~15, no.~2, pp. 1456--1468, Feb
  2016.

\bibitem{Abedi2016}
M.~R. Abedi, N.~Mokari, M.~R. Javan, and H.~Yanikomeroglu, ``Limited rate
  feedback scheme for resource allocation in secure relay-assisted ofdma
  networks,'' \emph{IEEE Transactions on Wireless Communications}, vol.~15,
  no.~4, pp. 2604--2618, April 2016.

\bibitem{Javan2016}
M.~Javan, N.~Mokari, F.~Alavi, and A.~Rahmati, ``Resource allocation in
  decode-and-forward cooperative communications networks with limited rate
  feedback channel,'' \emph{IEEE Transactions on Vehicular Technology},
  vol.~PP, no.~99, pp. 1--1, 2016.

\bibitem{Yang2016}
X.~Yang and A.~L. Swindlehurst, ``Limited rate feedback in a mimo wiretap
  channel with a cooperative jammer,'' \emph{IEEE Transactions on Signal
  Processing}, vol.~64, no.~18, pp. 4695--4706, Sept 2016.

\bibitem{Wang2016a}
L.~Wang, Y.~Cai, Y.~Zou, W.~Yang, and L.~Hanzo, ``Joint relay and jammer
  selection improves the physical layer security in the face of csi feedback
  delays,'' \emph{IEEE Transactions on Vehicular Technology}, vol.~65, no.~8,
  pp. 6259--6274, Aug 2016.

\bibitem{Eltayeb2014}
M.~E. Eltayeb, T.~Y. Al-Naffouri, and H.~R. Bahrami, ``Compressive sensing for
  feedback reduction in mimo broadcast channels,'' \emph{IEEE Transactions on
  Communications}, vol.~62, no.~9, pp. 3209--3222, Sept 2014.

\bibitem{Lv2016}
Z.~Lv and Y.~Li, ``A channel state information feedback algorithm for massive
  mimo systems,'' \emph{IEEE Communications Letters}, vol.~20, no.~7, pp.
  1461--1464, July 2016.

\bibitem{Kuo2012}
P.~H. Kuo, H.~T. Kung, and P.~A. Ting, ``Compressive sensing based channel
  feedback protocols for spatially-correlated massive antenna arrays,'' in
  \emph{2012 IEEE Wireless Communications and Networking Conference (WCNC)},
  April 2012, pp. 492--497.

\bibitem{Huang2013}
P.~Huang, M.~Tonnemacher, Y.~Du, D.~Rajan, and J.~Camp, ``Towards scalable
  network emulation: Channel accuracy versus implementation resources,'' in
  \emph{INFOCOM, 2013 Proceedings IEEE}, April 2013, pp. 1959--1967.

\bibitem{Huang2013a}
P.~Huang, D.~Rajan, and J.~Camp, ``Weibull and suzuki fading channel generator
  design to reduce hardware resources,'' in \emph{Wireless Communications and
  Networking Conference (WCNC), 2013 IEEE}, April 2013, pp. 3443--3448.

\bibitem{Huang2016}
P.~Huang, Y.~Du, and Y.~Li, ``Stability analysis and hardware resource
  optimization in channel emulator design,'' \emph{IEEE Transactions on
  Circuits and Systems I: Regular Papers}, vol.~63, no.~7, pp. 1089--1100, July
  2016.

\bibitem{Huang2013c}
P.~Huang, D.~Rajan, and J.~Camp, ``An autoregressive doppler spread estimator
  for fading channels,'' \emph{IEEE Wireless Communications Letters}, vol.~2,
  no.~6, pp. 655--658, 2013.

\bibitem{Zhou2013}
M.~Zhou, L.~Zhang, L.~Song, and M.~Debbah, ``A differential feedback scheme
  exploiting the temporal and spectral correlation,'' \emph{IEEE Transactions
  on Vehicular Technology}, vol.~62, no.~9, pp. 4701--4707, Nov 2013.

\bibitem{Jeon2016}
Y.~S. Jeon, H.~M. Kim, Y.~S. Cho, and G.~H. Im, ``Time-domain differential
  feedback for massive miso-ofdm systems in correlated channels,'' \emph{IEEE
  Transactions on Communications}, vol.~64, no.~2, pp. 630--642, Feb 2016.

\bibitem{Kim2012}
K.~Kim, T.~Kim, D.~J. Love, and I.~H. Kim, ``Differential feedback in
  codebook-based multiuser mimo systems in slowly varying channels,''
  \emph{IEEE Transactions on Communications}, vol.~60, no.~2, pp. 578--588,
  February 2012.

\bibitem{Zhang2012}
L.~Zhang, L.~Song, M.~Ma, and B.~Jiao, ``On the minimum differential feedback
  for time-correlated mimo rayleigh block-fading channels,'' \emph{IEEE
  Transactions on Communications}, vol.~60, no.~2, pp. 411--420, February 2012.

\bibitem{Nevat2016}
I.~Nevat, G.~W. Peters, K.~Avnit, F.~Septier, and L.~Clavier, ``Location of
  things: Geospatial tagging for iot using time-of-arrival,'' \emph{IEEE
  Transactions on Signal and Information Processing over Networks}, vol.~2,
  no.~2, pp. 174--185, June 2016.

\bibitem{Szurley2017}
J.~Szurley, A.~Bertrand, and M.~Moonen, ``Topology-independent distributed
  adaptive node-specific signal estimation in wireless sensor networks,''
  \emph{IEEE Transactions on Signal and Information Processing over Networks},
  vol.~3, no.~1, pp. 130--144, March 2017.

\bibitem{Jake1974}
W.~Jake, \emph{Microwave Mobile Communication}.\hskip 1em plus 0.5em minus
  0.4em\relax Piscataway, NJ: Wiley-IEEE Press, 1974.

\bibitem{Huang2017}
P.~Huang, W.~Wang, and Y.~Pi, ``Estimation on channel state feedback overhead
  lower bound with consideration in compression scheme and feedback period,''
  \emph{IEEE Transactions on Communications}, vol.~65, no.~3, pp. 1219--1233,
  March 2017.

\bibitem{Huang2014a}
P.~Huang and D.~Rajan, ``Bounds on the overhead of spectrum sensing in
  cognitive radio,'' in \emph{2014 IEEE Global Communications Conference}, Dec
  2014, pp. 846--850.

\bibitem{Huang2014b}
\BIBentryALTinterwordspacing
P.~Huang and Y.~Pi, ``An improved location service scheme in urban environments
  with the combination of {GPS} and mobile stations,'' \emph{Wireless
  Communications and Mobile Computing}, vol.~14, no.~13, pp. 1287--1301, 2014.
  [Online]. Available: \url{http://dx.doi.org/10.1002/wcm.2232}
\BIBentrySTDinterwordspacing

\bibitem{Huang2017a}
P.~Huang, ``Study on a low complexity ecg compression scheme with multiple
  sensors,'' \emph{arXiv preprint arXiv:1704.01612}, 2017.

\bibitem{Sayed2003}
A.~H. Sayed, \emph{Fundamentals of Adaptive Filtering}.\hskip 1em plus 0.5em
  minus 0.4em\relax New York, NY, USA: Wiley, 2003.

\bibitem{Berger1975}
T.~Berger, \emph{Rate Distortion Theory and Data Compression}.\hskip 1em plus
  0.5em minus 0.4em\relax Vienna: Springer Vienna, 1975, pp. 1--39.

\bibitem{Berger1971}
\BIBentryALTinterwordspacing
------, \emph{Rate Distortion Theory: A Mathematical Basis for Data
  Compression}, ser. Prentice-Hall electrical engineering series.\hskip 1em
  plus 0.5em minus 0.4em\relax Prentice-Hall, 1971. [Online]. Available:
  \url{https://books.google.com/books?id=-HV1QgAACAAJ}
\BIBentrySTDinterwordspacing

\bibitem{Zamir2008}
R.~Zamir, Y.~Kochman, and U.~Erez, \emph{IEEE Transactions on Information
  Theory}, no.~7, pp. 3354--3364, July.

\end{thebibliography}
\bibliographystyle{IEEEtran}

\end{document}